\documentclass{PoS}

\title{A Compact High Energy Camera (CHEC) for the Gamma-ray Cherenkov Telescope of the Cherenkov Telescope Array}
\ShortTitle{CHEC}

\author{R. White\\
	Max-Planck-Institut f\"{u}r Kernphysik, P.O. Box 103980, 69029 Heidelberg, Germany\\
	E-mail: \email{richard.white@mpi-hd.mpg.de}}

\author{\speaker{H. Schoorlemmer}\\
	Max-Planck-Institut f\"{u}r Kernphysik, P.O. Box 103980, 69029 Heidelberg, Germany\\
	E-mail: \email{harmscho@mpi-hd.mpg.de}}

\author{for the CTA GCT project\\
	 	http://www.cta-observatory.org}
        
\abstract{The Gamma-ray Cherenkov Telescope (GCT) is one of the Small Size Telescopes (SSTs) proposed for the Cherenkov Telescope Array (CTA) aimed at the 1 TeV to 300 TeV energy range. GCT will be equipped with a Compact High-Energy Camera (CHEC) containing 2048 pixels of physical size about 6$\times$6~mm$^2$, leading to a field of view of over 8 degrees. Electronics based on custom TARGET ASICs and FPGAs sample incoming signals at a gigasample per second and provide a flexible triggering scheme. Waveforms for every pixel in every event are read out are on demand without loss at over 600 events per second. A GCT prototype in Meudon, Paris saw first Cherenkov light from air showers in late 2015, using the first CHEC prototype, CHEC-M. This contribution presents results from lab and field tests with CHEC-M and the progress made to a robust camera design for deployment within CTA.}

\FullConference{35th International Cosmic Ray Conference -ICRC217-\\
		10-20 July, 2017\\
		Bexco, Busan, Korea}

\begin{document}

\section{Introduction}
\label{intro}

The Cherenkov Telescope Array (CTA) will host $\sim$70 Small-Sized Telescopes (SST)~\cite{sst-icrc} on the Southern Hemisphere site. The SSTs will provide CTA with sensitivity in an energy range from about 1 to 300 TeV and an angular resolution unmatched by any instrument above X-ray energies. The Gamma-ray Cherenkov Telescope (GCT) is one proposed option for the SSTs~\cite{gct-gamma16,gct-tel}. The GCT telescope is a dual mirror Schwarzschild-Couder design with a primary mirror of diameter 4~m, a secondary mirror of diameter 2~m, a focal length of 2.3~m resulting in a spherical focal plane with a radius of curvature 1.0~m~\cite{robast}. GCT will be equipped with the Compact High-Energy Camera (CHEC), also suitable for use in the SST design proposed by the ASTRI groups of CTA~\cite{astri}. CHEC is designed to record flashes of Cherenkov light lasting from a few to a few tens of nanoseconds, with typical image width and length of $\sim$0.2$^{\circ} \times $1.0$^{\circ}$ and promises a low-cost, high-reliability, high-data-quality solution for a dual-mirror SST. The small focal length of the telescope implies that an approximately 0.2$^{\circ}$ angular pixel size is achievable with pixels of physical dimensions of 6 to 7 mm (matched to the aformentioned mimimum image width), while the dual-mirror optics ensure that the point spread function (PSF) of the telescope is below 6~mm up to field angles of 4.5$^{\circ}$. A field of view (FoV) of 8$^{\circ}$ (required to capture high-energy and off-axis events) can therefore be covered with a camera of diameter about 0.4 m, composed of 2048 pixels. This allows the use of commercially available photosensor arrays, significantly reducing the complexity and cost of the camera. Multi-anode photomultipliers (MAPMs) and silicon photomultipliers (SiPMs) are under investigation via the development of two prototypes CHEC-M and CHEC-S respectively.

\section{Design}
\label{design}

\begin{figure}[ht]
  \centering
  \makebox[1\columnwidth][c]{\includegraphics[trim=3cm 2.5cm 2.3cm 1.3cm,
    clip=true, width=1\textwidth]{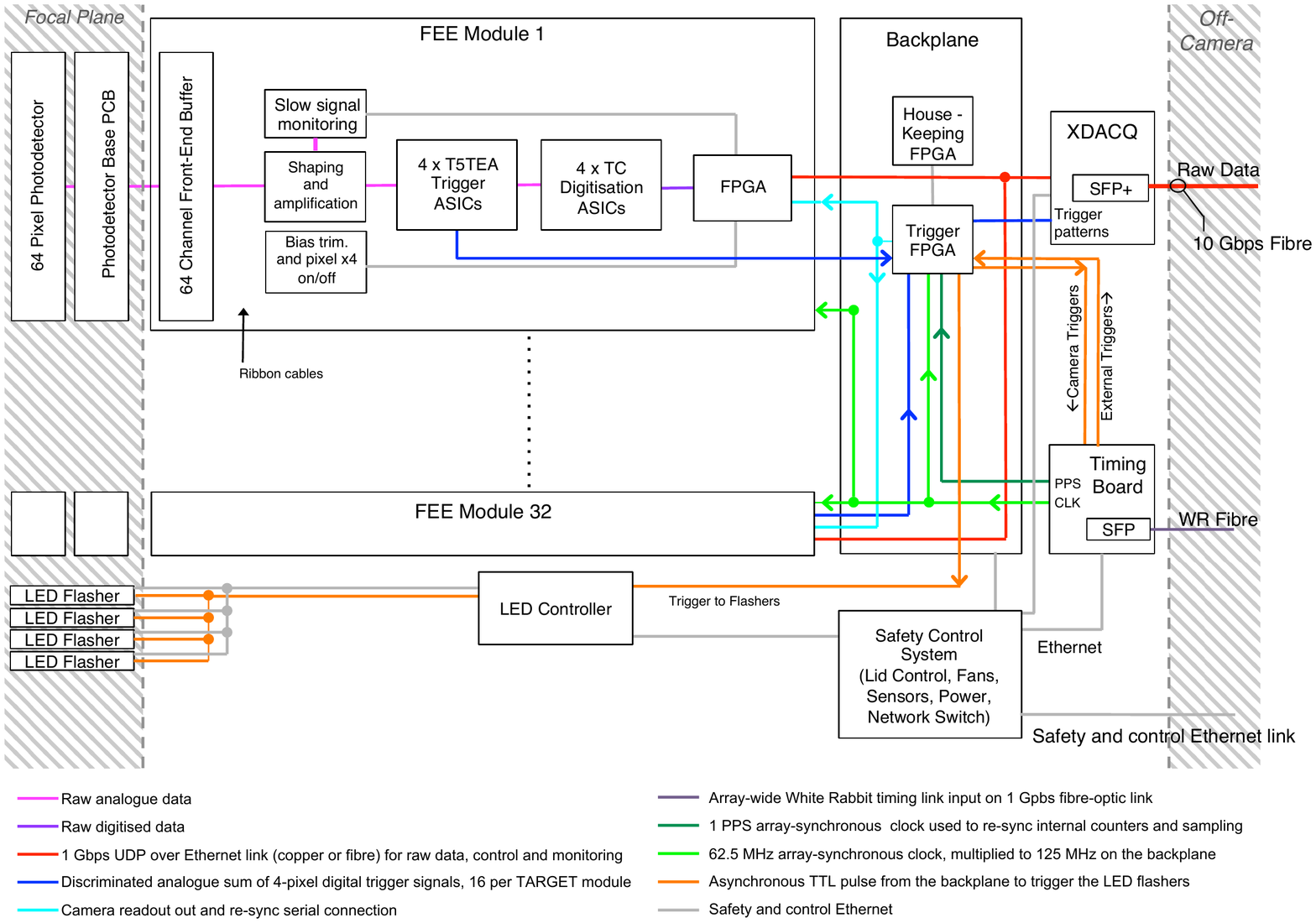}}
  \caption{A schematic showing the logical elements of CHEC,
    the communication between those elements, the raw data flow
    through the camera, the trigger architecture and the clock
    distribution scheme. Power distribution is excluded for
    simplicity.}
  \label{fig:arch}
\end{figure} 

\begin{figure}[ht]
  \centering
\resizebox{1\columnwidth}{!}{\includegraphics[trim=2cm 1cm 2cm 5.8cm, clip=true]{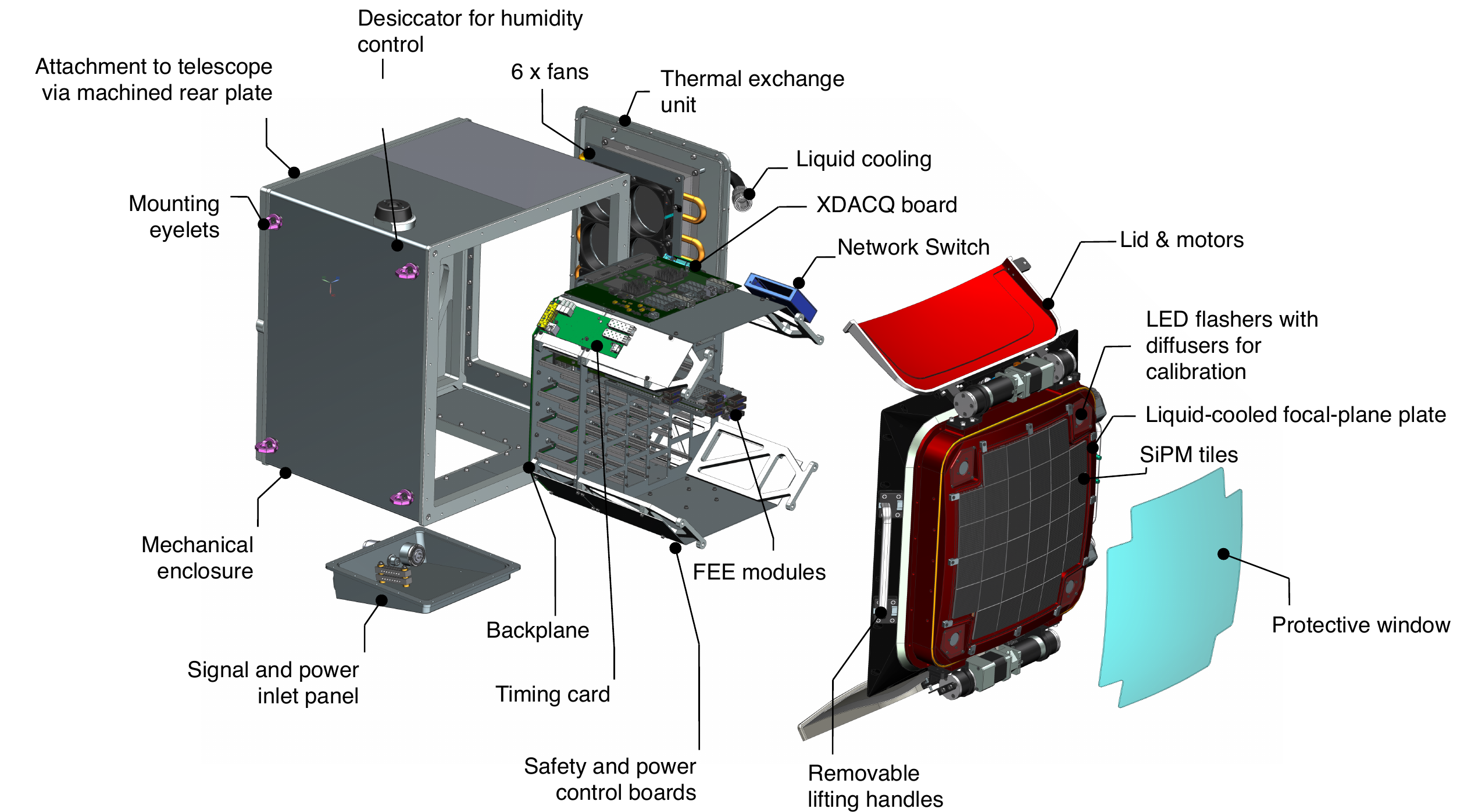}}
\caption{The CHEC-S CAD model with the key elements highlighted.}
  \label{fig:cam}
\end{figure} 

The architecture of the CHEC internal elements is shown in Figure \ref{fig:arch} whilst Figure~\ref{fig:cam} shows an exploded CAD image indicating the key camera components. The 2048 CHEC pixels are instrumented as 32 photosensors each comprising 64 pixels of $\sim$6$\times$6~mm$^2$ and arranged in the focal plane to approximate the required radius of curvature resulting from the telescope optics. CHEC-M utilises Hamamatsu H10966B MAPMs whereas CHEC-S is based around Hamamatsu S12642-1616PA-50 SiPM tiles (for further details of the CHEC-M design see~\cite{checm-icrc}). For CHEC-S each SiPM tile contains 256 $\sim$3$\times$3~mm$^2$ pixels that are combined in groups of four on bias board directly mounted to the SiPM to provide the desired camera pixel size.

\begin{figure}[ht]
  \centering
\resizebox{1\columnwidth}{!}{\includegraphics[trim=2cm 4.5cm 2cm 1cm, clip=true]{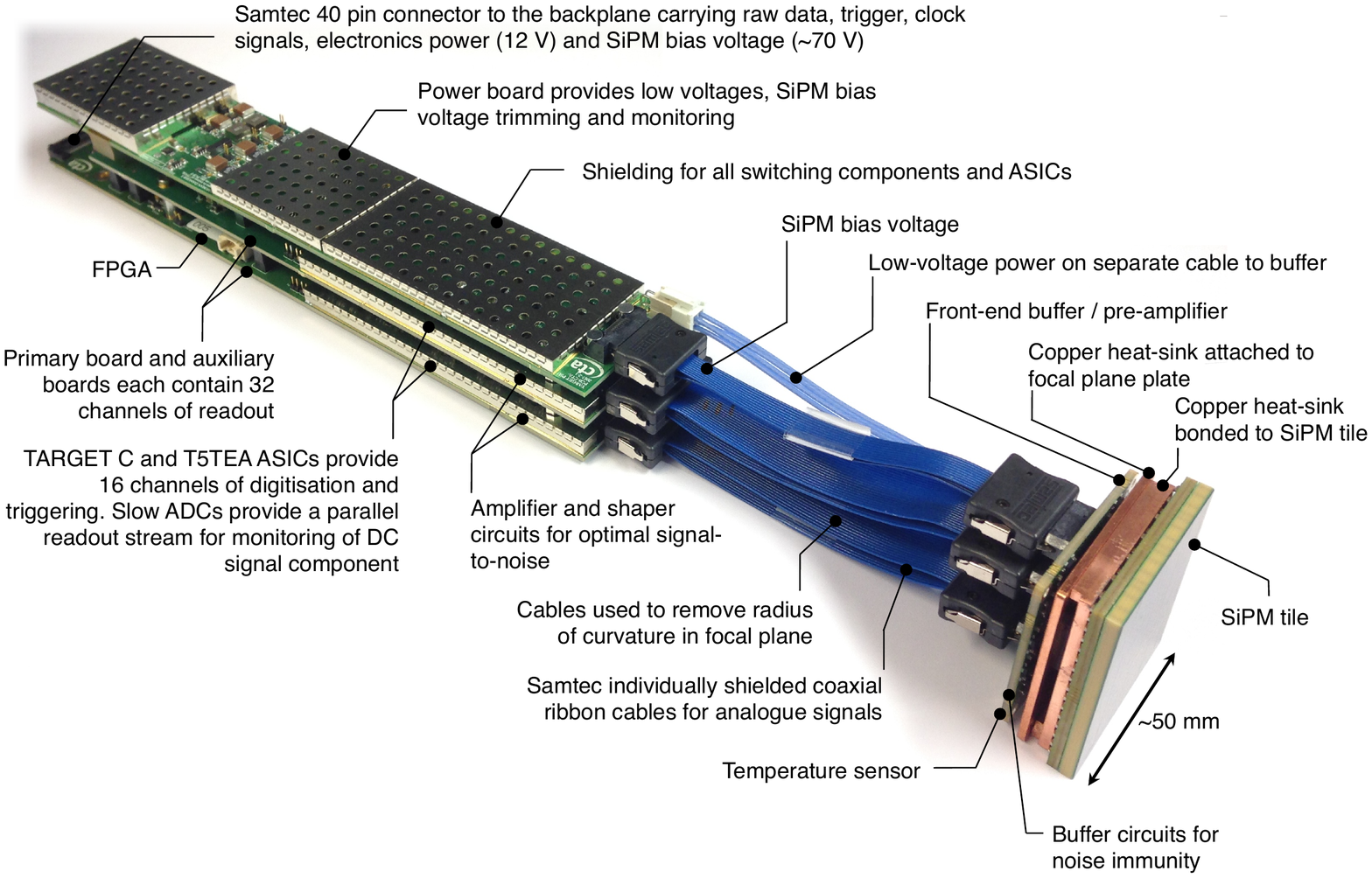}}
\caption{The CHEC-S SiPM tile and FEE module. The final FEE module is likely to be very similar to that shown, whereas the SiPM will see an upgrade to the latest technology (see Section~\ref{prospects}).}
  \label{fig:tm}
\end{figure}

Front-end electronics (FEE) modules, based on TARGET ASICs~\cite{target-gamma16}, connect to each photosensor. FEE modules provide full-waveform digitisation for every channel and the first-level of camera trigger as the discriminated analogue sum of four neighbouring pixels. Figure~\ref{fig:tm} shows the FEE module in detail. A pre-amplifier buffer board connects directly to the photosensor to provide noise immunity for signal transport to the sampling and trigger boards. Individually shielded ribbon cables further minimise the influence of noise and remove the curvature at the focal plane, allowing the use of a planar internal rack to house the modules. 

Simulations show that the optimal pulse width for triggering is around 5 to 10~ns \footnote{Narrower and the time gradient of Cherenkov images across neighbouring pixels forming the analogue sum prevents pile-up to reach the trigger threshold, wider and night-sky-background (NSB) photons limit the performance of the analogue sum trigger.}. To achieve this in CHEC-M the MAPM pulses are widened via the preamplifer directly behind the MAPM. The SiPM pulses in CHEC-S require shortening, which is achieved via a zero-pole shaping circuit on the FEE modules. 

The CHEC-M FEE modules are based on TARGET 5 ASICs, 16-channel devices combining digitisation and triggering functionalities~\cite{t5}. Since CHEC-M, TARGET has undergone development over several generations of ASIC. The latest FEE modules employ four 16-channel TARGET C ASICs for sampling and four 16-channel T5TEA ASICs for triggering. The TARGET C ASIC is a 12-bit device, which when used within CHEC provides an effective dynamic range of 1 to $\sim$500~pe (with the recovery of larger signals off line possible due to the waveform digitisation). The sampling rate is tunable, but will nominally be set to 1~GSa/s for CHEC. TARGET C contains a 64~ns deep analogue sampling array followed by a storage array with a maximum depth of 16384~ns. Each cell must be calibrated, and therefore to minimise the calibration needed for CHEC-S, the storage array is configured to 4096~ns. The position of the readout window digitised from storage array is selectable with nanosecond resolution with a size setable in 32~ns blocks, nominally set to 96~ns for CHEC (chosen to capture high-energy, off axis events as they transit the focal plane). Each TARGET module includes a slow-signal digitisation chain, providing a per-pixel measurement of the DC light level in the photosensors that may be used to track the pointing of the telescope via stars during normal operation. The TARGET module accepts a 12~V input for all electronics use and $\sim$70~V for the SiPM bias voltage, which is then trimmed to a precise value per four camera pixels on-board. An FPGA on-board each FEE module is used to configure the ASICs and other module components, to read-out raw data from the ASICs, and to package and buffer raw data for output from the module. Module control and raw data output is via UDP over a 1~Gbps Ethernet link at the rear of the modules. 


A backplane provides the power, clock, trigger and data interface the FEE modules. The backplane forms a nanosecond accurate camera trigger decision by combing trigger signals from all FEE modules in a single FPGA. The trigger FPGA, a Xilinx Virtex 6, accepts all 512 first-level trigger lines from the FEE modules and implements a camera-level trigger algorithm (currently requiring a coincidence between two neighbouring FEE trigger patches). Upon a positive camera trigger, a message is sent to the FEE modules to retrieve data from the sampling ASICs at the appropriate position in their memory. Data links to the FEE modules are routed via the backplane to a data-acquisition board(s) (DACQ) and routed off-camera via fibre-optic link. In CHEC-M two prototype DACQ boards were used to provide four 1~Gbps links, whilst CHEC-S and beyond will see the use of the a new board providing a single 10~Gbps out of the camera. A timing board provides absolute timing to the camera via an array-wide White Rabbit system. A safety board intelligently controls power to camera components based on monitored environmental conditions. 

LED flasher units placed in the corner of the camera provide calibration over a range of illumination intensities via reflection from the secondary mirror~\cite{led-icrc2015}. An external lid system provides protection from the elements. The camera consumes $\sim$450~W and is powered by a single supply, providing 12~V and 70~V DC, mounted at the rear of the secondary mirror. Thermal control of the camera is via an external chiller mounted on the telescope. Chilled liquid is circulated through the camera focal plane plate (via hollow ribs) and a thermal exchange unit on the camera body. Six fans internal to the camera circulate the resulting cooled air. Cooling the focal plane plate allows the SiPM temperature to be maintained at a level desirable to stabilise the gain. The camera is hermetically sealed and a breather-desiccator is used to maintain an acceptable level of humidity and instrument neutral pressure.

\section{Prototype Commissioning and Performance}
\label{proto}

\subsection{CHEC-M}
\label{checm}

CHEC-M and the components therein have undergone extensive lab testing, here a brief summary is given. The overall angular-averaged\footnote{The dual-mirror optical system results in light from angles of up to 70$^\circ$ impinging on the focal plane.} detection efficiency of a CHEC-M pixel, including camera dead space is around 14.5\%. The single photoelectron (pe) peak can be resolved for all camera pixels when operated at the manufacturer's highest recommended voltage of 1100~V. During normal operation the MAPMs are operated at a lower gain, and extrapolation is required from single pe measurements to gain-match the camera. Each MAPM accepts only a single HV supply for all 64 pixels, and incurs an irreducible spread in gain of $\sim$20\% RMS. Whilst the gain of individual pixels may be calibrated off-line for event reconstruction, on-line CHEC-M may not be gain matched to better than this spread, an important factor in the performance of the camera trigger. 

\begin{figure}[t]
  \centering
\resizebox{1\columnwidth}{!}{\includegraphics[trim=2cm 1.8cm 2cm 6.8cm, clip=true]{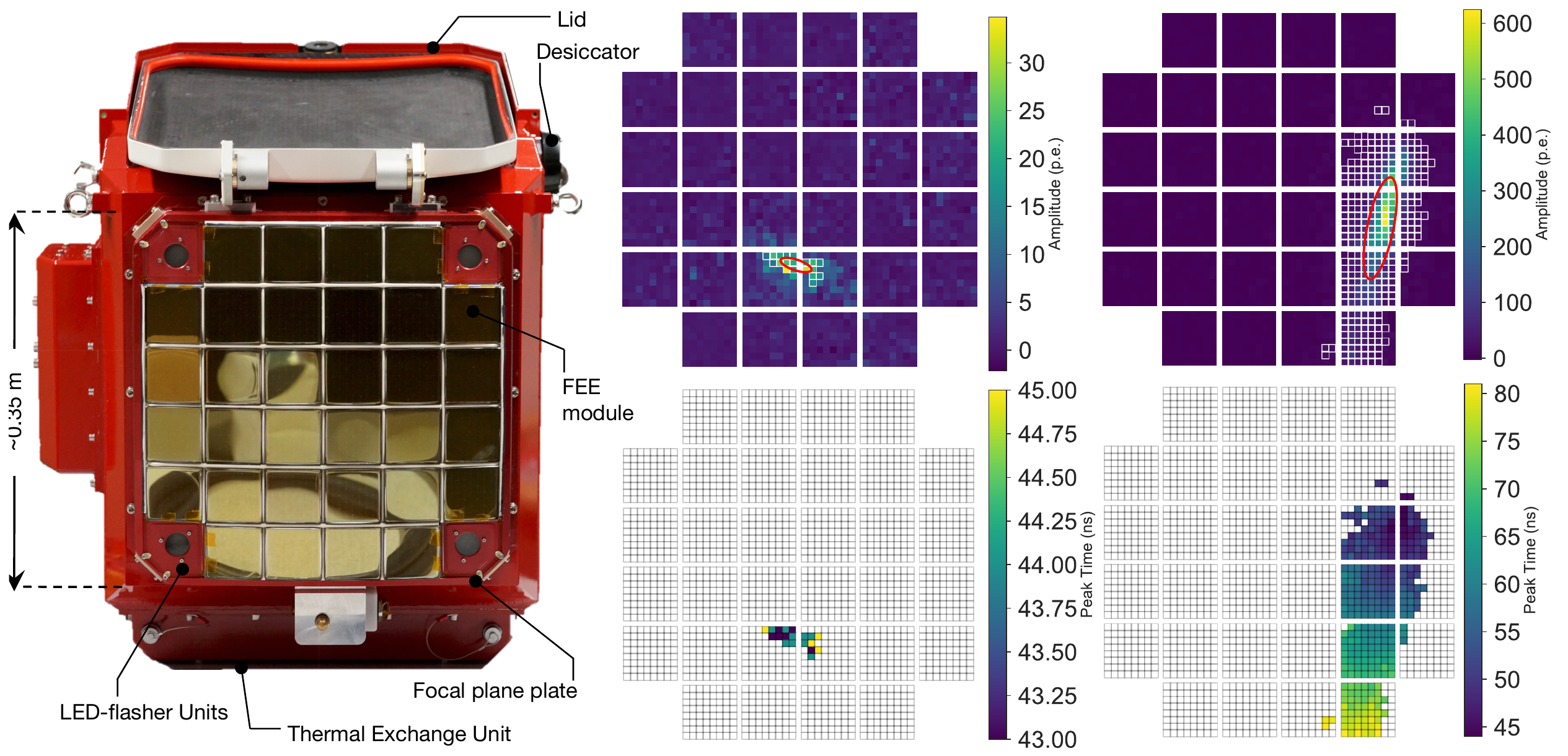}}
\caption{A photograph of the CHEC-M prototype camera (left) and examples of two Cherenkov images recorded with CHEC-M on the GCT telescope prototype in Meudon, Paris (right). The intensity in each pixel is shown in the upper plots, whilst the peak arrival time per pixel is shown for the same images in the lower plots. For further explanation refer to the text.}
  \label{fig:res1}
\end{figure} 

A single pe pulse from the MAPM, with typical amplitude of 0.8~mV is translated by the pre-amplifier to a pulse of typical peak voltage 2.4~mV and FWHM 5.5$\pm$1~ns as measured across all 2048 pixels. The maximum pre-amplifier output pulse height of $\sim$1.2~V is matched to the maximum input voltage of the TARGET 5 ASICs and corresponds to $\sim$500~pe. The output of the H10966 is linear to within 20\% at 1000~pe, and the dynamic range of CHEC-M is therefore limited by the digitisation range of the TARGET 5 ASICs. Charge resolution above saturation is possible by fitting the saturated waveforms. One ADC count in CHEC-M corresponds to approximately 0.3~mV or 0.13~pe, quantisation errors are therefore always much smaller than poisson fluctuations. The electronic noise of the FEE module system is $\sim$1~mV (or $\sim$0.5~pe) RMS. Timing measurements indicate a pulse arrival time of $\pm$0.9~ns relative to the camera trigger, within the CTA requirement of 1~ns. The TARGET 5 ASICs combine sampling, digitisation and analogue triggering in the same package. Coupling between sampling and triggering operations limit the CHEC-M trigger sensitivity to $\sim$5~pe per pixel depending on the gain of the pixel in question. Simulations indicate that the desired trigger threshold per analogue of four pixels is $\sim$10~pe, corresponding to a minimum of $\sim$2.5~pe per pixel. Figure~\ref{fig:res1} shows the fully assembled CHEC-M prototype camera to the left. 

\begin{figure}[ht]
  \centering
\resizebox{1\columnwidth}{!}{\includegraphics[trim=2cm 2cm 2cm 7cm, clip=true]{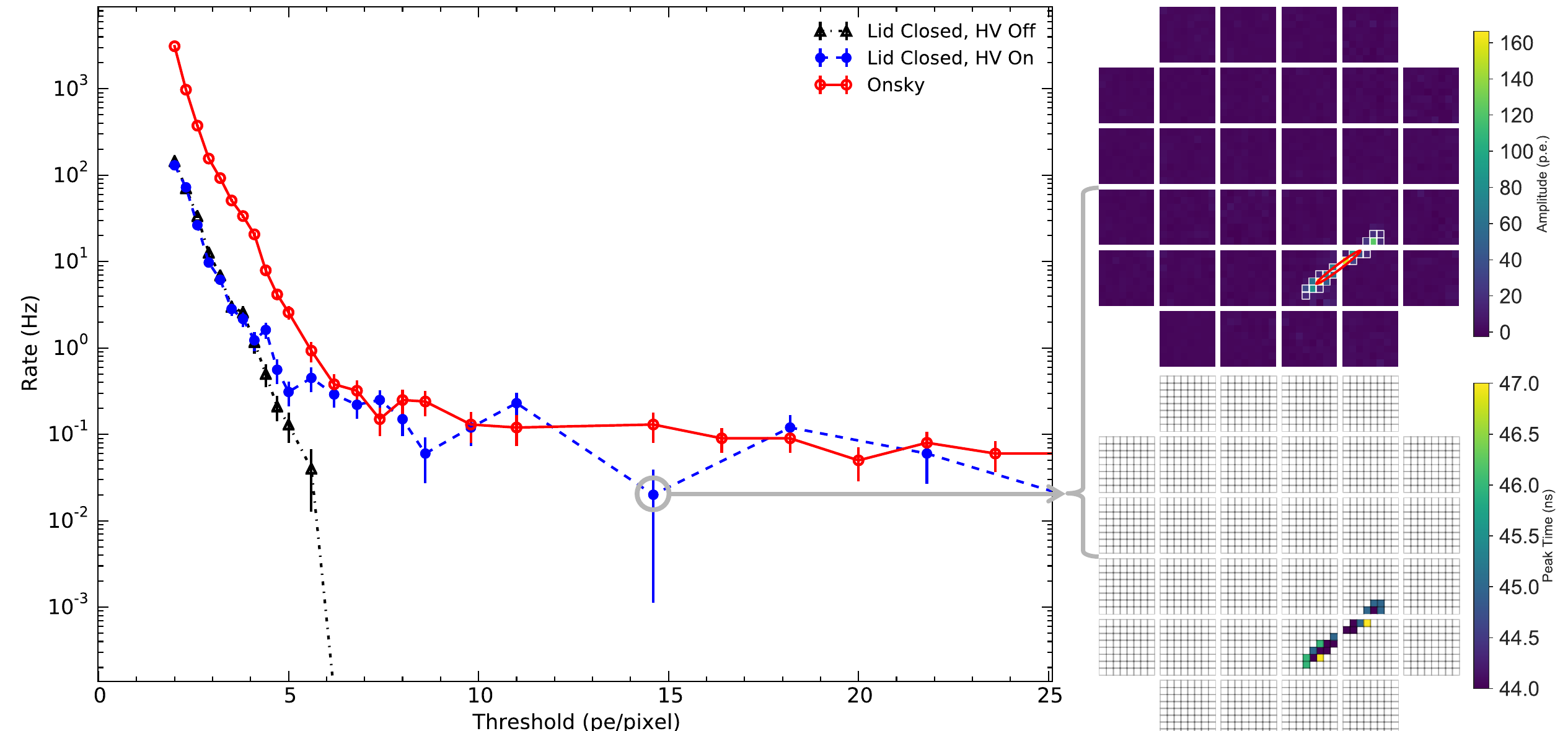}}
\caption{Trigger rate as a function of camera-trigger threshold as recorded for CHEC-M on the GCT telescope prototype in Meudon, Paris. The rate recorded with the camera lid closed and the MAPM HV off is indicated by the black triangles (broken-dashed line). The rate with the MAPM HV on is shown by the blue filled circles (dashed line). The rate recorded on-sky at an elevation angle of 60$^\circ$ is then shown by the red open circles (solid line). Inset to the right an image recorded with the HV on and the lid closed is shown (intensity in each pixel to the top and peak arrival time per pixel to the bottom).}
  \label{fig:res2}
\end{figure} 

CHEC-M has been deployed on the GCT prototype structure at the Observatoire de Paris in Meudon near Paris for two intensive observing campaigns. During the first campaign in November 2015, the first Cherenkov light by a CTA prototype, and a dual-mirror telescope, was recorded~\cite{jason-gamma16}. Figure~\ref{fig:res1} (right) shows two examples of on-sky Cherenkov images are shown from the second campaign in Spring 2017. The upper most images show the calibrated image intensity in pe for each camera pixel. Observations in Meudon took place under a NSB light level estimated to be 20 to 100 times brighter than at the actual CTA site. Such observing conditions required operating CHEC-M at a low gain, particularly far from the 1100~V point at which the previously mentioned absolute gain calibration is possible resulting in a an RMS spread in the calibrated gain between all pixels of $\sim$15\%. The white boxes outlining part of the images indicate pixels that survive image cleaning, in which all pixels containing with a signal greater than 20~pe and at least one neighbouring pixel with a signal larger than 10~pe, or vice versa, are retained. Ellipses resulting from the extracted Hillas parameters are shown in red. For further images and a comparison to Monte Carlo see~\cite{gct}. The lower images show the peak arrival time for all camera pixels for the same events as the upper images. As expected for a Cherenkov flash from a shower inclined with respect to the telescope focal plane the images can be seen to propagate across the focal plane in time. This additional information is only possible due to the waveform sampling nature of the camera electronics, and will be useful for advanced image cleaning, background rejection and event reconstruction algorithms. Additionally images at the highest energies can take many tens of nanoseconds to cross the camera, as can be seen for the right-most Cherenkov image in Figure~\ref{fig:res1}, without a $\sim$100~ns read out window such images would appear truncated, negatively impacting the analysis. The images shown here, and indeed all those recorded to date with CHEC-M, are attributed to air-showers stemming from cosmic-rays rather than gamma rays. 

To determine an appropriate operating threshold whilst on site, a scan of trigger rate as a function of threshold was taken (see Figure~\ref{fig:res2}). In the image shown the TARGET 5 sampling was disabled to avoid the previously mentioned problem of coupling between sampling and triggering operations. In black, data taken with the MAPM HV off is shown, indicating the level of electronic noise in the system. Turning the HV on with the camera lid closed results in the blue curve. A steady rate of events at the 0.1~Hz level are observed in the telescope park position (0$^\circ$ elevation) above a threshold of roughly 5~pe, attributable to cosmic rays interacting directly with the camera. Such an event is shown inset to the right of Figure~\ref{fig:res2}. On-sky data with the telescope at 60$^\circ$ elevation is shown in red. Accidental triggers due to fluctuations in the NSB dominate at low thresholds. Above $\sim$5~pe the rate is dominated by triggers from Cherenkov light initiated by cosmic-rays, as seen in the right of Figure~\ref{fig:res1}. Roughly 10$\%$ of these triggers were characteristic of cosmic rays interacting directly with the camera (the reduction in rate from that seen with the lid closed is due to the geometric effect of elevating the camera). The unique geometry and fast time profile of these events make them easy isolate. During normal operations in Meudon the trigger threshold was set to $\sim$11~pe pixel resulting in a steady Cherenkov event rate of $\sim$0.1~Hz. 

The prototyping of CHEC-M has proven invaluable in the development of a reliable, high-performace, product for CTA. Deployment in the field helped to verify interfaces and to improve the planned operations procedure. Regular operation has proven critical in understanding system stability and reliability. On-sky data is proving useful in the development of the data-analysis chain and in understanding the levels of calibration that will be required for CTA. The BEE used in CHEC-M will be close to those of production-phase system, with triggering, clock-syncronisation and data-transmission from  2~k channels of electronics all working succesfully. However, CHEC-M does not meet all the CTA performance requirements, with the non-uniformity in gain and the trigger noise incurred with sampling enabled being of greatest concern.

\subsection{CHEC-S}
\label{checs}

The components for CHEC-S are currently under test with camera integration expected this year. CHEC-S will tackle the limiting factors in the CHEC-M performance. The use of SiPMs allows gain measurements easily for a range of input illumination levels and bias voltages, and may even be determined in the absence of light from the dark counts intrinsic to the SiPM. This will improve the off-line calibration and charge reconstruction (simulations show that better than 10\% RMS between all pixels is needed to meet CTA requirements). The SiPM gain spread is also intrinsically much less than that seen in MAPMs, and the bias voltage is adjustable per four camera pixels, so gain matching to much higher precision than in CHEC-M will be possible. The gain of SiPMs is temperature sensitive, and for CHEC-S will drop by approximately 10-20\% over a 10$^\circ$~C increase. The liquid cooled focal plane plate will stabilise the temperature to within $\pm$1$^\circ$~C over time scales for which the gain may easily be re-measured in-situ. The angle-average detection efficiency for CHEC-S will also be significantly higher than for CHEC-M, not least due to less dead space in the focal plane. Due to the different wavelength dependence of the SiPM response, NSB rates are higher than in the MAPM case, even if radiation at wavelengths $>$550~nm is blocked before the camera. This background increase should be compensated by the improvement in efficiency for signal photons. Dark count rates from the SiPMs at the nominal operating gain and temperature have been measured to be less than $\sim20\%$ of the expected dark sky NSB rate, ensuring a negligible impact on performance. Due to the undesirable coupling between sampling and triggering in the TARGET 5 ASICs, functionalities were split into two separate ASICs. T5TEA provides triggering based on the same concept as TARGET 5, with a sensitivity reaching the single pe level and a trigger noise of 0.25~pe for the CHEC-S gain. TARGET C performs sampling and digitisation, with a $\sim$70\% larger dynamic range and with an improvement in charge resolution by a factor $>$2 with respect to TARGET 5~\cite{target-gamma16}. 

\section{Future Prospects}
\label{prospects}

Once prototyping is complete we plan to construct and deploy three CHEC cameras on the Southern-Hemisphere CTA site during a `pre-production' phase. During the production phase of CTA we aim to provide cameras for a significant fraction of the 70 baseline SSTs. It is expected that the majority of components used in CHEC-S will also be used in the final production design of CHEC with the exception of the photosensors. SiPM technology is rapidly evolving and the latest devices offer significant performance improvements compared to the Hamamatsu S12642 used in CHEC-S, including increased photo-detection efficiency, lower optical cross talk and a reduced dependency of gain on temperature~\cite{sipm}. Additionally there may be performance advantage associated with an enlarged field of view that may be obtained by using 7~mm rather than 6~mm pixels. Laboratory tests of the latest SiPMs and simulations with different pixel sizes are ongoing, with the aim of choosing a photosensor for the pre-production CHEC cameras this year. 

\acknowledgments
\label{ack}
We gratefully acknowledge financial support from the agencies and organizations listed here: http://www.cta-observatory.org/consortium\_acknowledgments. This work was conducted in the context of the CTA GCT project.

\end{document}